\documentclass[preprint,showpacs,preprintnumbers,amsmath,amssymb,
nofootinbib,eqsecnum,12pt,floatfix]{revtex4}

\usepackage{color}
\usepackage{amsmath}
\usepackage{graphicx}

\def\drawbox#1#2{\hrule height#2pt
        \hbox{\vrule width#2pt height#1pt \kern#1pt
              \vrule width#2pt}
              \hrule height#2pt}

\def\Asym#1#2{\vcenter{\vbox{\drawbox{#1}{#2}
              \kern-#2pt 
              \drawbox{#1}{#2}}}}

\newlength{\dummysp}
\newcommand{\beq}{\begin{eqnarray}}
\newcommand{\eeq}{\end{eqnarray}}

\newcommand{\p}{{\partial}}

\newcommand{\e}{{\epsilon}}
\newcommand{\s}{{\sigma}}
\newcommand{\vev}[1]{{\langle #1 \rangle}}

\newcommand{\ord}[1]{{{\cal O}(#1)}}
\newcommand{\gappeq}{\mathrel{\rlap {\raise.5ex\hbox{$>$}}
{\lower.5ex\hbox{$\sim$}}}}
\newcommand{\lappeq}{\mathrel{\rlap{\raise.5ex\hbox{$<$}}
{\lower.5ex\hbox{$\sim$}}}}
\newcommand{\myref}[1]{(\ref{#1})}

\newcommand{\ket}[1]{{ | #1 \rangle }}
\newcommand{\bra}[1]{{ \langle #1 | }}

\newcommand{\ben}{\begin{enumerate}}
\newcommand{\een}{\end{enumerate}}

\newcommand{\psib}{{\bar \psi}}

\newcommand{\bit}{\begin{itemize}}
\newcommand{\eit}{\end{itemize}}

\newcommand{\qbf}{\boldsymbol{q}}

\newcommand{\tfirst}{t_{\text{first}}}

\def\[{\left [}
\def\]{\right ]}
\def\({\left (}
\def\){\right )}

\def\nott#1{\setbox0=\hbox{$#1$}                
   \dimen0=\wd0                                 
   \setbox1=\hbox{/} \dimen1=\wd1               
   \ifdim\dimen0>\dimen1                        
      \rlap{\hbox to \dimen0{\hfil/\hfil}}      
      #1                                        
   \else                                        
      \rlap{\hbox to \dimen1{\hfil$#1$\hfil}}   
      /                                         
   \fi}                                         %

\begin{document}

\title{Probes of nearly conformal behavior in lattice
simulations of minimal walking technicolor}

\author{$^{\color{blue}{\clubsuit}}$Simon {\sc Catterall}}
\email{smc@physics.syr.edu}
\author{$^{\color{blue}{\natural}}$Joel {\sc Giedt}}
\email{giedtj@rpi.edu}
\author{$^{\color{blue}{\heartsuit}}$Francesco {\sc Sannino}}
\email{sannino@cp3.sdu.dk}
\author{$^{\color{blue}{\clubsuit}}$Joe {\sc Schneible}}
\email{jschneib@physics.syr.edu}
\affiliation{\vspace{10pt} \\
$^{\color{blue}{\clubsuit}}$Department of Physics, Syracuse University, NY 13244.\vspace{5pt} \\ 
$^{\color{blue}{\natural}}$Department of Physics,
Applied Physics and Astronomy, 110 8th St., Rensselaer Polytechnic Institute, Troy, 
NY 12180 USA.\vspace{5pt} \\ 
$^{\color{blue}{\heartsuit}}${ CP}$^{ \bf 3}${-Origins}, 
Campusvej 55, DK-5230 Odense M, Denmark.\footnote{{ C}entre 
of Excellence for { P}article { P}hysics { P}henomenology 
devoted to the understanding of the {Origins} of Mass in the universe.}}
\begin{flushright}
{\it CP$^3$- Orgins: 2009-14}
\end{flushright}

\date{Oct.~22, 2009}

\begin{abstract}
We present results from high precision, large volume simulations
of the lattice gauge theory corresponding to minimal walking
technicolor.  We find evidence that the pion decay constant
vanishes in the infinite volume limit and that the dependence of
the chiral condensate on quark mass $m_q$ is inconsistent 
with spontaneous symmetry breaking.  These findings are
consistent with the all-orders beta function prediction as well as the Schr\"odinger functional studies that
indicate the existence of a nontrivial infrared fixed point.
\end{abstract}


\maketitle

\section{Minimal Conformal Gauge Theories}
Depending on the number of flavors, matter representation and colors, 
non-abelian gauge theories are expected to exist in a 
number of distinct phases, classifiable according 
to the force felt between two static sources. 
The knowledge of this phase diagram is relevant 
for the construction of extensions of the 
Standard Model (SM) that invoke dynamical electroweak symmetry
breaking \cite{Weinberg:1979bn,Susskind:1978ms}.  
It is also useful in providing ultraviolet completions of 
unparticle models \cite{Georgi:2007ek,Sannino:2008nv,Sannino:2008ha}.  

``Minimal walking technicolor'' and similar models
employ fermions in higher dimensional representations 
of the new gauge group \cite{Sannino:2004qp,Dietrich:2005jn,Dietrich:2006cm,
Foadi:2007ue,Ryttov:2008xe}.  It is thought that some of these 
theories will develop a non-trivial infrared fixed point (IRFP)
for a small number of flavors \cite{Sannino:2004qp,Ryttov:2007cx}.
The presence of a {\it bona fide} IRFP requires the vanishing of the beta
function for a certain value of the coupling.  However,
it may be possible (at least in perturbation theory) 
to find a scheme where the beta function 
has a zero yet no IRFP actually exists; indeed there
are known examples in supersymmetric theories, when
the beta function is written in 't Hooft's scheme \cite{tHoRG}.  
On the other hand,
if the beta function is written in a scheme that uniquely and correctly
determines scheme-independent quantities at the fixed point --- such 
as the anomalous dimension (scaling exponent) of the fermion mass 
operator --- then it is a ``physical'' beta function.  We discuss
such a beta function in this article --- the conjectured all-orders
beta function (cf.~Eq.~\myref{aobf} below).  It vanishes at $g=g_*$ such that
$\beta_0 - \frac{2}{3} T(r) N_f \gamma(g_*^2)=0$
where $\gamma(g^2)$ is the anomalous dimension of $\psib \psi$
and $T(r)$ is the Dynkin index of the representation $r$
and $N_f$ is the number of flavors.  Since the one-loop coeffient 
$\beta_0$ is universal,
it can be seen that in this scheme the vanishing of the
beta function leads to an unambiguous result for $\gamma(g_*^2)$,
which is physical.

Historically a nearly conformal behavior has been identified with the 
slow rise of the coupling constant, in an unspecified renormalization scheme, as
the energy scale is reduced.  Such a slow rise was termed {\it walking} 
behavior \cite{Eichten:1979ah,Holdom:1981rm,Yamawaki:1985zg,Appelquist:1986an}.
It can be shown that this is a scheme dependent statement (a
nice illustration was made in \cite{deldeb}).
We expect, however, that a large anomalous dimension in an
on-shell scheme is meaningful in the desired fashion --- it generates
a condensate that is large compared to the scale
``$\Lambda_{QCD}$'' of the theory.

In order to better establish the location of the conformal window in
minimal walking technicolor models there have recently 
been a number of lattice studies\footnote{Searches for the conformal
window in theories with fundamental representation
quarks have also received recent attention 
\cite{Appelquist:2009ty,Appelquist:2009ka,Fodor:2009wk,Fodor:2008hn,
Deuzeman:2009mh}}\cite{Catterall:2007yx,Catterall:2008qk,
Shamir:2008pb,DeGrand:2008kx,DelDebbio:2008wb,DelDebbio:2008zf,Hietanen:2008vc,
Hietanen:2009az,Pica:2009hc}. 
These investigations indicate that these gauge theories are nearly or
actually conformal, as predicted in \cite{Sannino:2004qp,Ryttov:2007cx}.
Conformality at large distances implies scale invariance,
which forbids a chiral condensate.  Stable lattice simulations 
require a small but nonzero fermion mass, which ensures that
the lattice theory will possess a condensate.
The condensate vanishes when the mass is extrapolated to zero
at finite volume, so lattice measurements of the condensate 
must be analyzed carefully to disentangle the 
infinite volume, zero mass continuum behavior from the effects of small
but non-zero quark masses and finite volumes.
In principle, we should be able to distinguish true scale
invariance from walking behavior via a careful study 
of the spectrum and chiral condensate.  The point is that the
two cases will show different behavior as the mass and inverse
volume are sent to zero, where the latter extrapolation should
be performed first.  This will be the aim of
the current study which focuses on the minimal walking technicolor
model. This is an $SU(2)$ gauge theory with two Dirac
flavors (four Weyl or Majorana fermions\footnote{Since
we add a mass term in the lattice formulation, it is the Majorana description
which is more appropriate.}) 
transforming according to the adjoint representation of the gauge group.  
The global quantum symmetry group is $SU(4)$.  A Majorana mass term
reduces this to $SO(4)$.  The lattice results that we report here
are a continuation
of our earlier work \cite{Catterall:2007yx,Catterall:2008qk}.
We will present results with spatial volumes of
$8^3, 12^3, 16^3$ and $24^3$, at several values of the quark
mass.  We are able to see trends in the finite size effects
that are quite intriguing.

In Section \ref{apmc} we start with a brief summary of the analytical
predictions, making use of the old and new approaches.  
In addition we discuss the effect of introducing a 
non-zero fermion mass in a gauge theory that
is otherwise conformal.  Section \ref{latt} reviews our lattice
simulation results.  This is followed by an interpretation in Section
\ref{inte}.  Some experiments with clover fermions are
summarized in Section \ref{clov}, and we find that this improvement
should allow us to explore the $\epsilon$ and $\delta$ regimes
in a future work.  We conclude with a discussion in Section \ref{disc}.

\section{Theoretical Considerations}
\label{apmc}
We discuss here $SU(N)$ gauge theory with $N_f$ Dirac fermions 
in the adjoint representation, and the critical number
of flavors $N_f^{cr}$ below which scale invariance is
broken.  Because the quarks and
gluons are in the same representation, it is reasonable
to assume that $N_f^{cr}$ is independent of the number 
of colors $N_c$ (this is certainly true in perturbation
theory).  

\subsection{Truncated Schwinger-Dyson} 
The first analysis of the phase diagram with fermions in higher
dimensional representations used truncated 
Schwinger-Dyson (SD) equations, with fermions in
the two-index symmetric or antisymmetric 
representation \cite{Sannino:2004qp}. 
For $N_c=2$, the two-index symmetric representation is the 
adjoint representation.  The generalization of the SD approach to any 
representation was carried out in \cite{Dietrich:2006cm}, yielding
for the adjoint the conformal window
\begin{equation}
2.075 \lessapprox N_f^{\rm SD} < \frac{11}{4}=2.75 .
\end{equation}
In terms of Weyl fermions the window becomes $4.15 \lessapprox {N_W}_f^{\rm SD} \leq 5.5$ 
where the upper limit is the number of flavors above
which asymptotic freedom is lost.  The lower limit corresponds 
to the point when the SD equation can no longer be trusted and 
the anomalous dimension of the mass term is close to unity.  Thus
a theory of five Weyl fermions in the adjoint representation
would appear to be in the conformal window, but one is uncertain
what really occurs for the case of four Weyl fermions --- equivalent
to $N_f=2$ Dirac fermions.  Our lattice study seeks to address
this question; however, we first describe another analytical
estimate.  We mention in passing that the approach developed in \cite{Appelquist:1999hr} 
provides no useful constraint for any theory with fermions in higher 
dimensional representations as shown in \cite{Sannino:2005sk,Sannino:2009aw}.

\subsection{All-orders beta function and anomalous dimension}
Specializing the recently conjectured ``all-orders beta function'' \cite{Ryttov:2007cx}
to fermions in the adjoint representation, the $\beta$-function reads:
\beq
\beta(g) &=&- \frac{g^3}{(4\pi)^2} \frac{\beta_0 - \frac{2}{3} N_c N_f \gamma(g^2)}
{1- \frac{g^2}{8\pi^2} N_c \left( 1+ \frac{2\beta_0'}{\beta_0} \right)} \ ,
\label{aobf}
\eeq
where $\beta_0' = N_c (1 -N_f)$ and $\beta_0=\frac{N_c}{3} (11 - 4 N_f)$ is the
one-loop coefficient.

The all-orders beta function satisfies a number of consistency checks.
(i) The (exact) super-Yang-Mills result is recovered for $N_f=1/2$.
(ii) It compares well with the running 
of the Yang-Mills coupling constant as determined by lattice gauge theory.
(iii) It provides predictions consistent with the SD approach
for a critical value of $\gamma^c = 1$.
(iv) The conformal window matches the one 
obtained by a conjectured dual gauge theory.

Item (iv) relates to recent exact solutions 
of the 't Hooft anomaly matching conditions
\cite{Sannino:2009qc,Sannino:2009me}. Further developments appeared in \cite{Antipin:2009wr}.  
Naturally \myref{aobf} reduces to the well-known two-loop beta function
one when expanding to $\ord{g^5}$.  We give it here since
we will compare to it in the discussion below:
\beq
\beta (g) = -\frac{\beta_0}{(4\pi)^2} g^3 - \frac{\beta_1}{(4\pi)^4} g^5 \ ,
\label{perturbative}
\eeq
with (scheme independent) adjoint representation coefficients
\beq
\beta_0 = \frac{N_c}{3} (11 - 4 N_f), \quad
\beta_1 = \frac{N_c^2}{3} (34 - 32 N_f).
\label{12cof}
\eeq
We will also make use of
the anomalous dimension $\gamma =-{d\ln m}/{d\ln \mu}$ of the
renormalized mass $m$ to second order
\beq
\gamma = \frac{6 N_c g^2}{(4 \pi)^2} 
+ \frac{2 N_c(53 N_c - 5 N_f)g^4}{3(4\pi)^4} + {\cal O}(g^6)
\label{gammapert}
\eeq

The all-orders beta function predicts the anomalous dimensions 
of the fermion mass at the infrared fixed point 
and is in this sense ``physical.'' 
In \cite{Ryttov:2007cx} it was argued that the size of the conformal window
is determined by the largest value allowed for the 
anomalous dimension, $\gamma_c$:
\begin{equation}
\frac{11}{2(2+\gamma^c)} \leq N_f^{\rm BF} < \frac{11}{4}.
\label{secw}
\end{equation}
The fixed point value is
\begin{equation}
\gamma_* = \frac{11 - 4N_f}{2N_f} \ .
\end{equation}
(It is interesting to note that if we take $N_f=2$,
then $\gamma_*=3/4$ and
the lower end of the conformal window in
\myref{secw} is exactly $N_f=2$.)
If we use the SD inspired condition $\gamma^c = 1$ we would have
\begin{equation}
1.8\overline{3} \leq N_f^{\rm BF} < \frac{11}{4}
\end{equation}
whereas the maximal conformal window is achieved
in the unitarity limit $\gamma^c=2$: 
\begin{equation}
1.375 \leq N_f^{\rm BF} < \frac{11}{4}.
\end{equation}
What is important to notice is that
independently of which of these $\gamma^c$ is chosen, 
the prediction \cite{Sannino:2008ha,Ryttov:2007cx} 
is that the adjoint theory with 
$N_f=2$ has an IRFP and an associated anomalous 
dimension $\gamma = 3/4$.   Our aim here is
to scrutinize this prediction of the all-orders
beta function using lattice techniques. 

\subsection{Large anomalous dimensions at weak coupling} 
\begin{figure}
\begin{center}
\includegraphics[width=10cm ,height=7.0cm]{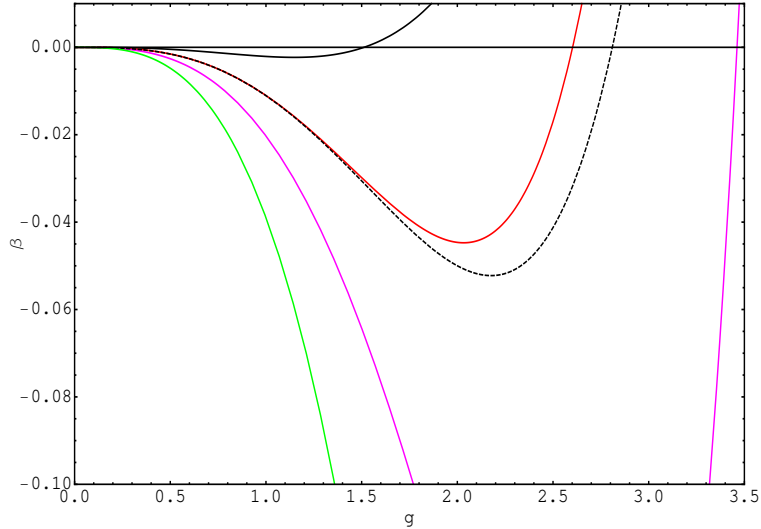}
\caption{Beta functions for different values of the number
of Dirac flavors in the adjoint representation 
of the $SU(2)$ gauge group. The black solid curve 
corresponds to $N_f=2.5$, the red to $N_f=2$, the 
dashed one is the two-loop beta function for $N_f=2$ again,
while the magenta curve corresponds to $N_f=1.5$. The green
curve is the beta function for super Yang-Mills.  \label{Beta-AD}}
\end{center}
\end{figure}

Often, in the literature, one finds plotted a
cartoon of the running of the coupling constant
for either conformal or nearly conformal theories. 
Here we provide yet another cartoon of this running 
but this time using the ``physical'' form of the
conjectured all-orders beta function \cite{Ryttov:2007cx}.
In order to be explicit, we augment this with a simplifying
ansatz for the dependence of the anomalous dimension
on the coupling constant, Eq.~\eqref{gammapert}.    
The advantage is that we will be able to plot what
happens when changing the number of flavors --- but this is
only a ``cartoon'' since we do not actually know
what $\gamma(g^2)$ really is in the true theory.
Thus, suppose we use the two-loop expression of the anomalous dimension 
together with the all-orders beta function.  Then the
beta function for various values of $N_f$ are shown in Fig.~\ref{Beta-AD}. 
We have also plotted the two-loop beta function 
for $SU(2)$ with $N_f=2$.  Interestingly the fixed point
is reached before the one from the two-loop beta function.  (This is 
consistent with the recent lattice results obtained in \cite{Hietanen:2009az}.)

What one should note from Fig.~\ref{Beta-AD} is that
a relatively small value of $g_*^2/(16 \pi^2)$ is obtained
while $\gamma_*=\ord{1}$.  Visually, this is because there
is a long renormalization group trajectory that must be
traversed in going from a $g \approx 0$ weak coupling
value to the $g_*$ fixed point.  The curve deepens as
the number of flavors is decreased from $N_f=2.5, 2, 1.5$,
consistent with the ordering of the fixed point values
$\gamma_* = 1/5, 3/4, 5/3$.
The general message is that one can have large values
of the anomalous dimensions and yet have coupling
constants at the IRFP which are small.

\subsection{The chiral condensate}
We have measured the chiral condensate through
the GMOR relation:
\beq
(m_\pi f_\pi)^2 = -2 m_q \Sigma.
\label{gmor}
\eeq
Here $\Sigma=\vev{\psib \psi}$ is the condensate
in infinite volume.  The GMOR relation
just follows from chiral symmetry
breaking with a small source $m_q$ for the ``scalar
current'' $\psib \psi$.  In the case of an IRFP $\Sigma$ must also
vanish as $m_q\to 0$.  Different scenarios for how it vanishes
have been discussed in \cite{Sannino:2008pz}.  The generic expectation
is that in theories where the anomalous dimension $\gamma <1$ or
theories where instanton effects are important such as the model
analyzed here, $\Sigma\sim m_q\Lambda_U^2$ with $\Lambda_U$ a high
energy scale characterizing the onset of asymptotic freedom.
In contrast QCD-like theories with chiral symmetry breaking 
possess a non-vanishing  condensate as $m_q\to 0$ in infinite
volumes.  

\subsection{Finite volume effects}
Lattice simulations are necessarily performed on a finite
four-dimensional volume, which we will denote $L^3 \times T$,
associating $T$ with the extent of the temporal
dimension.  If the theory possesses an IRFP,
then in the chiral limit $m_q \to 0$, large finite-size
effects will always be present.

It is well-established that there are three regimes possible for lattice
gauge theories with spontaneous chiral
symmetry breaking; the p-regime where $m_\pi L \gg 1$ and $m_\pi T \gg 1$,
the $\epsilon$-regime where $m_\pi L \ll 1$ and $m_\pi T \ll 1$ and the
$\delta$-regime with $m_\pi L \ll 1$ but $m_\pi T \gg 1$. 
In the small volume $\delta$ or $\epsilon$ regime
the chiral condensate will typically scale to zero linearly
with quark mass
in a manner similar to that expected in a theory with a IRFP.
Thus we must be particularly careful in interpreting our lattice
results on small boxes in order to distinguish the two scenarios.

Indeed, it 
is not clear that this categorization will prove useful in a
theory that has an IRFP, where there is no spontaneous chiral
symmetry breaking.  The difficulty is that the expansion
parameters and mode decoupling arguments rely heavily on
$f_\pi \not= 0$ in the chiral, infinite volume limit.  This
is not true for the theory with an IRFP.

\section{Lattice analysis}
\label{latt}
For details of our lattice action and simulation algorithm we refer
the reader to \cite{Catterall:2008qk}. Suffice it to say that we
have employed unimproved Wilson fermions in the
adjoint representation and a simple Wilson plaquette
action for the gauge field and generated configurations using
the usual HMC algorithm. We now
turn to the extraction of accurate estimates of the
meson and quark masses and the pion decay constant $f_\pi$.
\subsection{Current quark mass extraction}
The fermion mass $m_q$ is obtained from a fit to
\beq
G_{PCAC}(t) =
\frac{\p_t G_{AP}(t)}{G_{PP}(t)} \approx 
\frac{2 Z_m Z_P m_q }{Z_A} \equiv 2 m_{\text{PCAC}} , \quad 0 \ll t \ll T.
\label{ttft}
\eeq
Here, $T$ is the number of sites in the temporal direction,
$m_{\text{PCAC}}$  is the bare PCAC mass and $m_q$ is the
renormalized current quark mass.  
In this work we do not determine
the renormalization constants $Z_m, Z_P, Z_A$; 
however they are expected to be
$\ord{1}$ and we will suppress them in much
of what follows.  The two Green's functions involved in \myref{ttft} are:
\beq
G_{PP}^{ab}(t) = \int d^3x ~\vev{P^a(t,{\bf x}) P^b(0,{\bf 0})}, \quad
G_{AP}^{ab}(t) = \int d^3x ~ \vev{A_0^a(t,{\bf x}) P^b(0,{\bf 0})},
\label{Gdfs}
\eeq
where $P^a = \psib \gamma_5 t^a \psi$ and $A_0^a = \psib \gamma_0 \gamma_5 t^a \psi$,
with $t^a \in \{ \s^+, \s^-, \s^3 \}$.  For brevity we suppress the isospin indices $a,b$ and
leave it as implied that the nonvanishing components $G^{+-}$ (i.e. the ones
without disconnected diagrams) of the Green's functions
are used in the measurements.  At leading order in the expansion of states,
\beq
G_{PP}(t) \sim \p_t G_{AP}(t) \sim \cosh \[ m_\pi \( \frac{T}{2} -t  \) \]
 , \quad 0 \ll t \ll T.
\label{gppf}
\eeq
This is why a constant is expected in \myref{ttft}.
The integral over ${\bf x}$ in Eq.~\myref{Gdfs} projects onto zero momentum states.        
In practice we approximate
$\p_t \approx \nabla_t^{(S)}$, the symmetric difference
operator.

\subsection{Meson masses and decay constants}
Next we describe how $m_\pi$ and $f_\pi$ are measured.
Referring to the correlation function $G_{PP}(t)$,
we work in the leading exponential approximation:
\beq
G_{PP}(t) = C_{PP} \cosh \( m_\pi \( \frac{T}{2} - t \) \)
\eeq
On the other hand, using the resolution of the
identity in terms of states,
\beq
G_{PP}(t) = \frac{1}{2 m_\pi} \left| \bra{0} P(0,{\bf 0}) \ket{\pi, \qbf=0} \right|^2
2 e^{-m_\pi T/2} \cosh \( m_\pi \( \frac{T}{2} - t \) \)
\eeq
neglecting excited state contributions.
The matrix element $\bra{0} P(0,{\bf 0}) \ket{\pi, \qbf=0}$
is well known:
\beq
\bra{0} P(0,{\bf 0}) \ket{\pi, \qbf=0} = \frac{m_\pi^2 f_\pi}{2 Z_m Z_p m_q}
\eeq
Thus we obtain:
\beq
f_\pi = \( \frac{C_{PP}}{m_\pi^3} \)^{1/2} 2 Z_m Z_P m_q e^{m_\pi T / 4}
\eeq
This is then combined with \myref{ttft} to obtain
$f_\pi^{\text{bare}} \equiv f_\pi/Z_A$.

\subsection{Results at $\beta=2.05$}
First we discuss results from simulations
at $\beta=2.05$ on a $L^3 \times 32$ lattice with periodic boundary
conditions imposed in all directions.  For $L=8,12,16$
a total of 10500 HMC trajectories were generated.
For $L=24$, a total of approximately 5000 HMC trajectories
were generated.  In all cases,
the first 200 were discarded as thermalization
and observables were obtained by averaging
from every tenth trajectory of the remaining
ensemble.  Errors were corrected for autocorrelations
in the data.  We performed nonlinear fits to \myref{gppf}.
We estimated the statistical uncertainty by the jackknife method,
fitting repeatedly with a block of data removed.
Jackknife block sizes of 10, 25, 50, 100 and 200
all gave consistent results, including
error estimates.  

\begin{figure}
\includegraphics[width=2.5in,height=4in,angle=90]{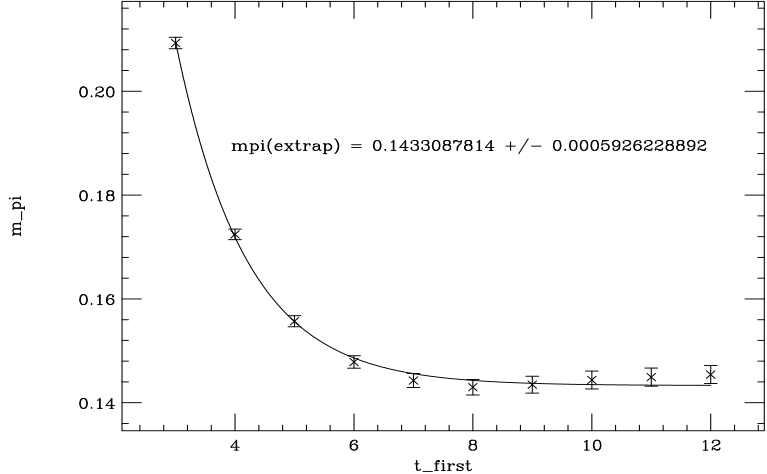}
\caption{Estimates of $m_\pi a$ for
$\beta=2.05$, $8^3 \times 32$ lattice as the beginning of
the fit range, $\tfirst$, is varied.
\label{fig3}}
\end{figure}

The fit results depend significantly on the
range of $t$ that is included, due to excited
state contamination.  The variable $\tfirst$
determines the first timeslice that is included.
An example of the fit variation is given
in Fig.~\ref{fig3}.  It can be seen that
there is a plateau that is reached as
$\tfirst$ is increased.  In practice we
fit to a constant plus exponential decay
with $\tfirst$, and use the constant with
fitting error as our estimate of a given
observable.

Results are shown in Tables
\ref{tab1} (mass variation) and \ref{tab1v} (volume variation).
The quantity $R$ will be discussed in a subsequent section.
It can be seen that $f_\pi$ decreases significantly
with volume, whereas the behavior of the pion and rho
masses are more complicated.
Further numerical analysis is presented in Table \ref{split},
which shows that a significant splitting of the
rho and pion occurs for large enough volume and small
enough quark mass.  This table also shows that we
are far from the heavy quark limit where $m_\pi \approx 2 m_q$
would hold.

\begin{table}
\begin{center}
\begin{tabular}{|c|c|c|c|c|c|c|}
\hline
$L$ & $ma$ & $m_\pi a$ & $m_\rho a$ & $f_\pi a$ & $m_q a$ & $R a^2$ \\
\hline
8 & -1.29 & 0.5686(9)  & 0.6010(6) & 0.619(7) & 0.1078(5) & 11.08(10) \\
8 & -1.30 & 0.3680(10) & 0.3952(12) & 0.668(10) & 0.0685(3) & 13.40(15) \\
8 & -1.31 & 0.1433(6)  & 0.1530(9) & 0.738(9) & 0.0253(2) & 19.55(8) \\ 
\hline
\end{tabular}
\caption{$\beta=2.05$, $L^3 \times 32$ PBC lattice, with
unimproved Wilson quarks. \label{tab1}
}
\end{center}
\end{table}

\begin{table}
\begin{center}
\begin{tabular}{|c|c|c|c|c|c|c|}
\hline
$L$ & $ma$ & $m_q a$ & $m_\pi a$ & $m_\rho a$ & $f_\pi a$  & $R a^2$ \\
\hline
8&	-1.31	& 0.0253(2)   &	0.1433(6)  &	0.1530(9)	 &  0.738(9)  &	19.55(8) \\
12&	-1.31	& 0.015236(63) &	0.1215(17) &	0.1547(24) &	0.4598(29)&	13.83(14) \\
16&	-1.31	& 0.01214(16) &	0.1075(15) &	0.1531(25) &	0.406(10) &	13.854(76) \\
24&	-1.31	& 0.00800(11) &	0.1254(42) &	0.1770(59) &	0.1743(46) &	8.25(25) \\
\hline
\end{tabular}
\caption{Quantities of interest for the
$\beta=2.05$, $L^3 \times 32$ PBC lattice, with
unimproved Wilson fermions. \label{tab1v}
}
\end{center}
\end{table}

\begin{table}
\begin{center}
\begin{tabular}{|c|c|c|c|}
\hline
$L$ & $m_q a$ & $m_\pi/m_q$ &	$(m_\rho-m_\pi)/m_\pi$ \\
\hline
8  &	0.0253(2)    & 5.664(51) &	0.0677(77) \\
12 &	0.015236(63) & 7.97(12)	 &  0.273(27) \\
16 &	0.01214(16)  & 8.86(17)	 &  0.424(31) \\
24 &	0.00800(11)  & 15.68(57) &	0.411(67) \\
\hline
\end{tabular}
\caption{Pion mass and rho-pion splitting enhancement
as $m_q$ and $1/L$ are decreased, for the
$\beta=2.05$, $L^3 \times 32$ PBC lattice, with
unimproved Wilson fermions. \label{split}
}
\end{center}
\end{table}

\subsection{Results at $\beta=2.5$}
Next we discuss results for $\beta=2.5$.  Here
we have also allowed for a larger time extent, $T=64$,
in order to account for what might be a finer
lattice spacing.  The fits versus $\tfirst$ are similar
to Fig.~\ref{fig3}.  The method of simulation, sampling and fits
are the same as for $\beta=2.05$.  Results are given
in Table \ref{tab2}.

\begin{table}
\begin{center}
\begin{tabular}{|c|c|c|c|c|c|c|}
\hline
$L$ & $ma$ & $m_\pi a$ & $m_\rho a$ & $f_\pi a$ & $m_q a$ & $R a^2$ \\
\hline
8  & -1.1 & 0.13625(7) & 0.14531(5) & 1.039(12) & 0.03834(3)  & 13.621(15) \\
12 & -1.1 & 0.12260(7) & 0.13537(15) & 0.593(11) & 0.02900(11) & 6.091(6) \\
16 & -1.1 & 0.1204(2)  & 0.1251(7) & 0.405(5)  & 0.0284(4)   & 2.957(13)  \\
24 & -1.1 & 0.1344(12) & 0.1497(6) & 0.242(2) & 0.0266(3) & 1.49(3) \\
\hline
\end{tabular}
\caption{Quantities of interest for the
$\beta=2.5$, $L^3 \times 64$ PBC lattice, with
unimproved Wilson fermions. \label{tab2}
}
\end{center}
\end{table}

\section{Interpretation}
\label{inte}
In this section we characterize the numerical
results.  First consider the quantity $R \equiv (m_\pi f_\pi/ m_q)^2=\Sigma/m_q$.
By our earlier arguments, for a theory with an IRFP,
this should approach a constant in the chiral limit $m_q \to 0$.
In a QCD-like theory, the order of limits (chiral versus thermodynamic)
matters.  If $L \to \infty$ before taking the chiral limit,
then $R$ would diverge inversely with the quark
mass $m_q$.  In the $\delta$- or $\e$-regime one has instead $\Sigma \sim m_q$,
which would lead to a finite result for $R$ in the chiral
limit.  Our values of $m_\pi L$ at the pseudo-critical
values of bare mass $ma$ range from 1.1 to 3.0, whereas in
the $\delta$- or $\e$-regime one has $m_\pi L \ll 1$.
Thus on our larger lattices we are certainly outside of
these small volume regimes.  In Tables~\ref{tab1v}
and \ref{tab2} we observe $R$ decreasing as $L$ increases.
This stands in stark contrast to
what would happen in a QCD-like theory that is
outside of the small volume regimes.  For this reason we
find that our data favors the IRFP interpretation,
as far as the chiral condensate is concerned.
In fact, we find that $R \sim 1/L^2$, as a result of
the observed scaling $f_\pi \sim 1/L$.
This seems to be evidence for a vanishing condensate
at large $L$, consistent with the existence of an IRFP. 
However, we cannot rule out a small but nonvanishing
infinite volume condensate; i.e., it is possible that we could just be seeing
a decreasing finite volume effect that is much larger
than the infinite volume piece for the lattices we
are studying.

\begin{figure}
\begin{center}
\includegraphics[width=5in,height=3in]{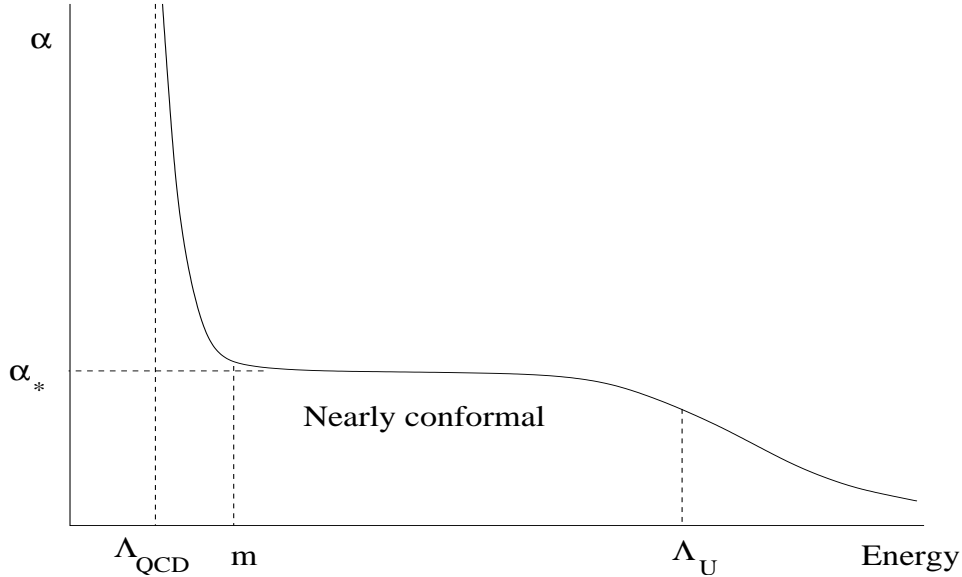}
\caption{Nearly conformal flow for a theory that
would have an IRFP when $m=0$.  The dynamical
scale ``$\Lambda_{QCD}$'' is generated below the
mass scale $m$.
\label{nclq}}
\end{center}
\end{figure}

The vanishing of the decay constant $f_\pi$ with increasing lattice
volume contrasts starkly with the approximate volume
independence of the pion (and rho) masses shown in Tables~\ref{tab1v} and \ref{tab2}.
The origin of this behavior is difficult to understand.
However, we can assume that at distances scales longer than
the inverse quark mass the fermions effectively decouple from
the dynamics leaving an IR theory which behaves like quenched
QCD but with a light scale
\beq
\Lambda_{QCD} \sim m_q e^{-8\pi^2/g^2_*}
\eeq
(see Figure.~\ref{nclq}).
These light gluonic states have been seen in simulations
\cite{Pica:2009hc} at energies below that of the corresponding
pion and rho states.  
Notice also that while
the pion and rho masses do not scale with the
lattice volume they are not simply the sum of the two constituent
quark masses $m_q$, as emphasized in
Table \ref{split}.  Instead it is best to think of the
pion as being composed of two ``dressed'' quarks where the 
dressing represents the effects of these light gluonic degrees of
freedom.  A useful analogy would the $\phi$-system in QCD composed
of two strange quarks. 

We can parameterize the dependence of
the pion mass on the quark masses in a phenomenological way as
\beq
m_\pi= c m_q^{1-\tilde \delta} L^{-\tilde \delta} + \ord{a \Lambda_{\text{UV}}^2}.
\label{pias}
\eeq
However, this formula only explains the increase in
meson masses in going from $L=16$ to $L=24$ provided
the $\ord{a \Lambda_{\text{UV}}^2}$ effects become larger in this limit,
presumably due to larger renormalizations as the
number of degrees of freedom is increased.  Also,
in the presence of an IRFP there will be large cutoff
effects because one is explicitly breaking conformality,
an essential symmetry of the theory we are trying to study.
The pion is no longer a pseudo-Nambu-Goldstone boson,
due to the absence of spontaneous chiral symmetry breaking,
hence its mass is quite sensitive to this explicit
breaking of scale invariance.  For this reason one
expects large $\ord{a \Lambda_{\text{UV}}^2}$ effects, which
is precisely what we observe.
To be consistent with our numerical results
the (positive) exponent $\tilde \delta$ that
appears in \myref{pias} should be small.

Compatibility with the GMOR relation then implies that
\beq
f_\pi = c' m_q^{\tilde\delta} L^{\tilde\delta -1} + \ord{a \Lambda_{\text{UV}}^2}
\label{fpas}
\eeq
which has the merit of guaranteeing that $f_\pi$ vanishes both
with the lattice volume and also as $m_q\to 0$.  Because $f_\pi$
is taken from a ratio of the lattice derivative of a correlation
function to another correlation function, it is possible that
large $\ord{a \Lambda_{\text{UV}}^2}$ cutoff effects may be
absent, due to cancellations.  In fact, this would explain
our lattice data in the tables above, where $f_\pi$ is seen
to decrease both with $m_q$ and $1/L$; a large constant
term $\ord{a \Lambda_{\text{UV}}^2}$ does not seem to be
present.  Thus it may be that the leading lattice spacing
correction to $f_\pi$ is actually $\ord{a m_q^2}$.

We note that our $\beta=2.5$ data is roughly consistent with
Eqs.~\myref{pias} and \myref{fpas}, while the $\beta=2.05$
is less so.  For either value of $\beta$, the increase in the meson
masses in going from $L=16$ to $L=24$ is quite strange,
and seems on its face to be at odds with variational
arguments.\footnote{We thank R.~Brower for raising this point.
The essence of the argument is that on doubling the lattice
size, the original pion wavefunction can be periodically extended.
But one would expect that it is no longer the minimum energy
eigenstate in the pseudo-scalar channel, since new basis
states for a variational analysis are allowed on the 
larger lattice.  It follows that the pion mass
in the larger volume will be lower.}  
However, topological features could play
a nontrivial role in such finite volume considerations,
in a way that might resolve the apparent paradox.

\section{Valence clover on unimproved Wilson sea}
\label{clov}
In order to proceed to light quark masses and move into
the $\delta$- and $\epsilon$-regimes we have experimented with
simulations that utilize a clover-improved Wilson-Dirac propagator.
We have computed the pion mass on the same unimproved dynamical Wilson
configurations described in Section \ref{latt}.  Setting the coefficient of the 
clover term $c_{SW}=\ord{1}$, we have been able to achieve
$m_\pi a \lappeq 0.05$ by tuning the valence
quark mass.  This indicates that dynamical
clover fermions would allow for an exploration
of the $\delta$- and $\epsilon$-regimes, something that we plan 
to do in a future work.

\section{Discussion}
\label{disc}
We have presented results for the low lying meson masses, decay constants
and chiral condensate from simulations of the minimal walking 
technicolor
theory corresponding to two flavors of adjoint Dirac fermions in
$SU(2)$ gauge theory.  Data at two couplings $\beta=2.05$ and $\beta=2.5$
and a range of lattice volumes $L^3\times 32(64)$, with $L=8,12,16,24$
were shown. Unimproved Wilson fermions and Wilson glue are used 
and ensembles of O(10000) configurations accumulated at each set of
parameter values.
 
We have shown how the GMOR relation may be used to compute the chiral
condensate and discussed the relationship between the condensate
measured on the lattice and its continuum cousin. 
Our results are consistent with the vanishing of the condensate in the
infinite volume limit and hence the existence of an infrared fixed point. 
The dependence of $f_\pi$ and the pion and rho masses
on both the quark mass and lattice volume are shown to also support the
presence of such a fixed point, though other interpretations
are possible.

\section*{Acknowledgements}
JG was supported by Rensselaer faculty development funds. 
The work of S.C. is supported in part by DOE grant DE-FG02-85ER40237. 
All simulations were performed using facilities at RPI's Computational
Center for Nanotechnology and Innovation.


\begin{thebibliography}{99}

\bibitem{Weinberg:1979bn}
  S.~Weinberg,
  Phys.\ Rev.\  D {\bf 19}, 1277 (1979).

\bibitem{Susskind:1978ms}
  L.~Susskind,
  Phys.\ Rev.\  D {\bf 20}, 2619 (1979).

\bibitem{Georgi:2007ek}
  H.~Georgi,
  Phys.\ Rev.\ Lett.\  {\bf 98}, 221601 (2007)

\bibitem{Sannino:2008nv}
  F.~Sannino and R.~Zwicky,
  arXiv:0810.2686 [hep-ph].

\bibitem{Sannino:2008ha}
  F.~Sannino,
  arXiv:0804.0182 [hep-ph].

\bibitem{Sannino:2004qp}
  F.~Sannino and K.~Tuominen,
  Phys.\ Rev.\  D {\bf 71}, 051901 (2005)

\bibitem{Dietrich:2005jn}
  D.~D.~Dietrich, F.~Sannino and K.~Tuominen,
  Phys.\ Rev.\  D {\bf 72}, 055001 (2005)

\bibitem{Dietrich:2006cm}
  D.~D.~Dietrich and F.~Sannino,
  Phys.\ Rev.\  D {\bf 75}, 085018 (2007)

\bibitem{Foadi:2007ue}
  R.~Foadi, M.~T.~Frandsen, T.~A.~Ryttov and F.~Sannino,
  Phys.\ Rev.\  D {\bf 76}, 055005 (2007)

\bibitem{Ryttov:2008xe}
  T.~A.~Ryttov and F.~Sannino,
  arXiv:0809.0713 [hep-ph].

\bibitem{Ryttov:2007cx}
  T.~A.~Ryttov and F.~Sannino,
  Phys.\ Rev.\  D {\bf 78}, 065001 (2008)

\bibitem{tHoRG}
G.~'t Hooft, ``The renormalization group in quantum field theory,''
in {\it Under the spell of the gauge principle,}
World Scientific, Singapore, 1994.




\bibitem{Eichten:1979ah}
  E.~Eichten and K.~D.~Lane,
  Phys.\ Lett.\  B {\bf 90}, 125 (1980).

\bibitem{Holdom:1981rm}
  B.~Holdom,
  Phys.\ Rev.\  D {\bf 24}, 1441 (1981).

\bibitem{Yamawaki:1985zg}
  K.~Yamawaki, M.~Bando and K.~i.~Matumoto,
  Phys.\ Rev.\ Lett.\  {\bf 56}, 1335 (1986).

\bibitem{Appelquist:1986an}
  T.~W.~Appelquist, D.~Karabali and L.~C.~R.~Wijewardhana,
  Phys.\ Rev.\ Lett.\  {\bf 57}, 957 (1986).

\bibitem{deldeb}
L.~del Debbio, talk given at ``Universe in a Box,''
Lorentz Center, Leiden, August, 2009.


\bibitem{Catterall:2007yx}
  S.~Catterall and F.~Sannino,
  Phys.\ Rev.\  D {\bf 76} (2007) 034504.

\bibitem{Catterall:2008qk}
  S.~Catterall, J.~Giedt, F.~Sannino and J.~Schneible,
      JHEP {\bf 0811} (2008) 009.


\bibitem{Shamir:2008pb}
  Y.~Shamir, B.~Svetitsky and T.~DeGrand,
  Phys.\ Rev.\  D {\bf 78}, 031502 (2008).
  
\bibitem{DeGrand:2008kx}
  T.~DeGrand, Y.~Shamir and B.~Svetitsky,
  Phys.\ Rev.\  D {\bf 79}, 034501 (2009)
  [arXiv:0812.1427 [hep-lat]].
  
\bibitem{DelDebbio:2008wb}
  L.~Del Debbio, M.~T.~Frandsen, H.~Panagopoulos and F.~Sannino,
  JHEP {\bf 0806}, 007 (2008).

\bibitem{DelDebbio:2008zf}
  L.~Del Debbio, A.~Patella and C.~Pica,
  arXiv:0805.2058 [hep-lat].

\bibitem{Hietanen:2008vc}
  A.~Hietanen, J.~Rantaharju, K.~Rummukainen and K.~Tuominen,
  PoS {\bf LATTICE2008}, 065 (2008).

\bibitem{Hietanen:2009az}
  A.~J.~Hietanen, K.~Rummukainen and K.~Tuominen,
  arXiv:0904.0864 [hep-lat].

\bibitem{Pica:2009hc}
  C.~Pica, L.~Del Debbio, B.~Lucini, A.~Patella and A.~Rago,
  arXiv:0909.3178 [hep-lat].
  
\bibitem{Appelquist:2009ty}
  T.~Appelquist, G.~T.~Fleming and E.~T.~Neil,
  Phys.\ Rev.\  D {\bf 79}, 076010 (2009)
  [arXiv:0901.3766 [hep-ph]].

\bibitem{Appelquist:2009ka}
  T.~Appelquist {\it et al.},
  arXiv:0910.2224 [hep-ph].

  
\bibitem{Fodor:2009wk}
  Z.~Fodor, K.~Holland, J.~Kuti, D.~Nogradi and C.~Schroeder,
  arXiv:0907.4562 [hep-lat].

\bibitem{Fodor:2008hn}
  Z.~Fodor, K.~Holland, J.~Kuti, D.~Nogradi and C.~Schroeder,
  arXiv:0809.4890 [hep-lat].

\bibitem{Deuzeman:2009mh}
  A.~Deuzeman, M.~P.~Lombardo and E.~Pallante,
  arXiv:0904.4662 [hep-ph].


\bibitem{Appelquist:1999hr}
  T.~Appelquist, A.~G.~Cohen and M.~Schmaltz,
  Phys.\ Rev.\  D {\bf 60} (1999) 045003.
  
 \bibitem{Sannino:2005sk}
   F.~Sannino,
   Phys.\ Rev.\  D {\bf 72}, 125006 (2005).

 \bibitem{Sannino:2009aw}
   F.~Sannino,
   Phys.\ Rev.\  D {\bf 79}, 096007 (2009).

\bibitem{Sannino:2009qc}
  F.~Sannino,
  Phys.\ Rev.\  D {\bf 80}, 065011 (2009).

\bibitem{Sannino:2009me}
  F.~Sannino,
  arXiv:0909.4584 [hep-th].

\bibitem{Antipin:2009wr}
  O.~Antipin and K.~Tuominen,
  arXiv:0909.4879 [hep-ph].
  D.~D.~Dietrich,
  arXiv:0908.1364 [hep-th].
\bibitem{Sannino:2008pz}
  F.~Sannino,
  Phys.\ Rev.\  D {\bf 80}, 017901 (2009)
  [arXiv:0811.0616 [hep-ph]].

\end{thebibliography}
\end{document}